\documentclass[3p,times]{elsarticle}

\usepackage{ecrc}

\volume{00}

\firstpage{1}

\journalname{Nuclear Physics A}

\runauth{Kathrin Koch et al.}

\jid{nupha}

\usepackage{amssymb}

\begin{document}

\begin{frontmatter}



\dochead{}

\title{$\pi^{0}$ and $\eta$ measurement with photon conversions in ALICE in proton-proton collisions at $\sqrt{s}$ = 7 TeV}

%
\author{K.~Koch~for~the~ALICE~Collaboration}
\address{Physikalisches Institut, Ruprecht-Karls-Universit\"at Heidelberg, Heidelberg, Germany}

\ead{kathrin.koch@cern.ch}

\begin{abstract}
We present a measurement of the $\pi^{0}$ transverse momentum spectrum and of the $\eta$/$\pi^{0}$ ratio in proton-proton collisions at $\sqrt{s}$ = 7 TeV at the CERN LHC. In this analysis the reconstruction of $\pi^{0}$ and $\eta$ mesons has been done via photon conversions in the central tracking system of ALICE. Therefore, this method is completely independent from the electromagnetic calorimeters. It makes the $\pi^{0}$ ($\eta$) measurement possible down to $p_{t}$ = 0.4 (0.6) GeV/c with a very good resolution and a very small background. For ~$10^{8}$ pp collisions the $p_{t}$ reach is 7 GeV/c. The results are compared to NLO pQCD calculations. 
\end{abstract} 


\end{frontmatter}


\section{Introduction}
\label{sec:Introduction}
Since the beginning of 2010 the LHC has been delivering proton-proton collisions at $\sqrt{s}$ = 7 TeV, the highest energy ever achieved. The measurement of the $\pi^{0}$ and $\eta$ spectra provides a test of the applicability of pertubative QCD (pQCD) calculations at this energy. These spectra are also essential to extract direct photons from the large decay background coming mainly from $\pi^{0}$ and $\eta$ decays. Once the direct photon spectrum is extracted in heavy ion collisions, one can obtain information like temperature and pressure about the quark gluon plasma (QGP). In addition, the corrected  $\pi^{0}$ transverse momentum spectrum in proton-proton collisions will serve as a baseline to study a possible $\pi^{0}$ suppression in heavy ion collisions at LHC energies.
\section{Experimental setup}
\label{sec:ExperimentalSetup}
ALICE is a general purpose heavy ion experiment to investigate strongly interacting matter and the quark gluon plasma in nucleus-nucleus collisions at the CERN LHC. A detailed description of the ALICE setup is given in \cite{jinst}. In this analysis the Inner Tracking System (ITS) and the Time Projection Chamber (TPC) are used. The measurements via calorimeters can be found in \cite{calo}. In this analysis a fully calibrated data set of about $10^{8}$ minimum bias events recorded between March and June 2010 has been used. Details about the trigger conditions are explained in \cite{trigger}.
\section{Photon reconstruction}
\label{sec:PhotonReco}
\begin{figure}[t]
 	\begin{minipage}{7.5 cm}
		\centering
 		\includegraphics[height=0.2\textheight]{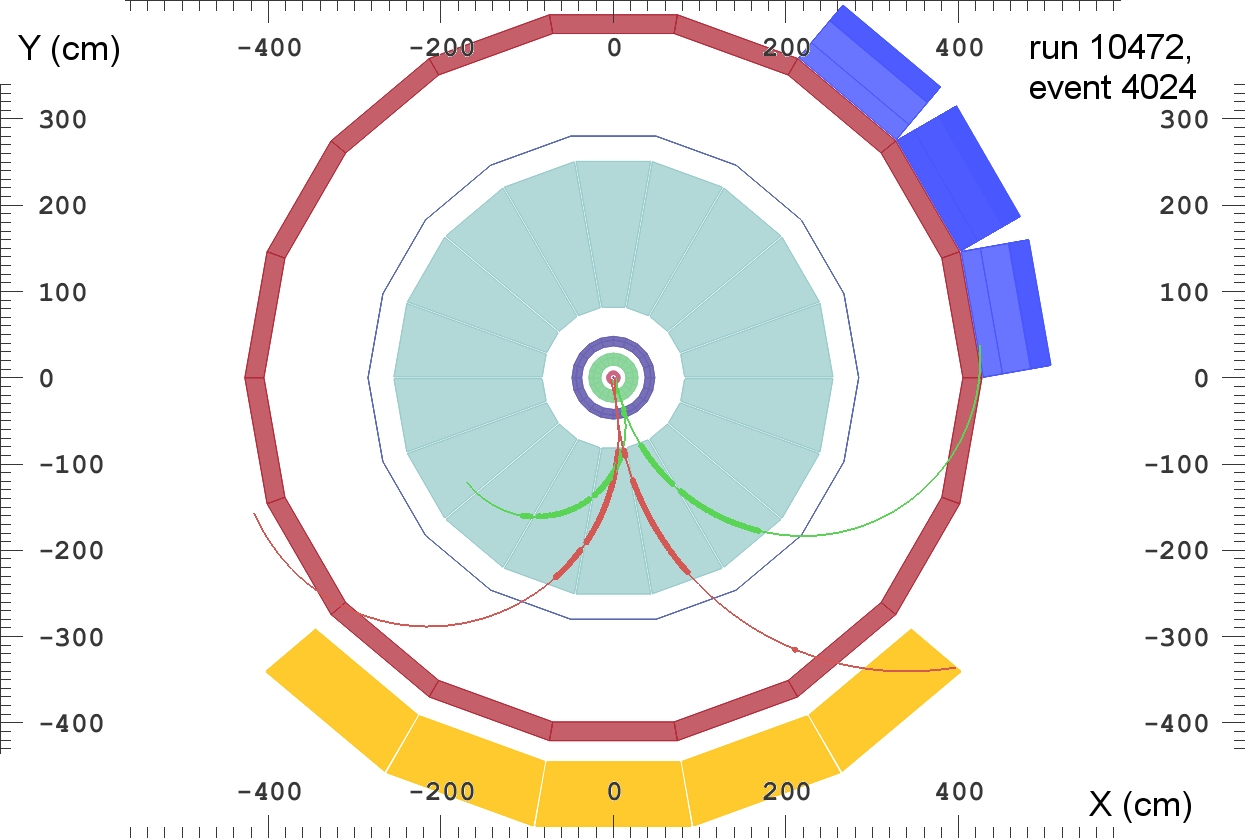} 
 	\end{minipage}
	\hspace{0.5cm}
	\begin{minipage}{7.5 cm}
  		\centering
	  	\includegraphics[height=0.25\textheight]{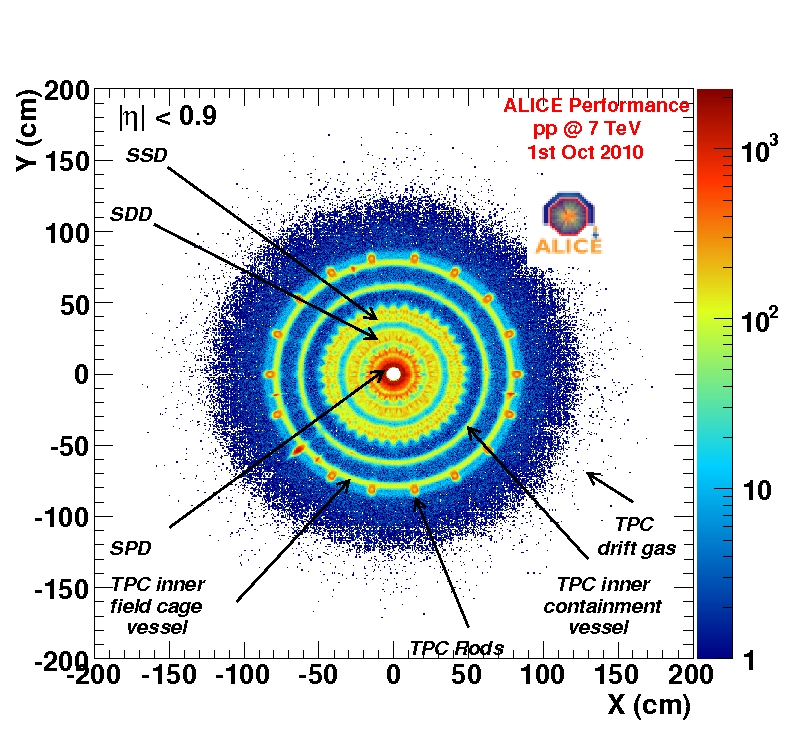}  
  	\end{minipage}
	\begin{minipage}{7.5 cm}
		\caption{Event display of a $\pi^{0}$ decay candidate. The red and green points are reconstructed clusters which are combined to a charged particle track. Here the two colors indicate the tracks belonging to two photons.}
		\label{fig:eventdisplay}	
 	\end{minipage}
	\hspace{0.5cm}
	\begin{minipage}{7.5 cm}
		\caption{Distribution of conversions in XY. The internal structure of ITS and TPC detectors is clearly visible.}
		\label{fig:XY}	
 	\end{minipage}
  	\begin{minipage}{7.5 cm}
		\centering
    		\includegraphics[height=0.2\textheight]{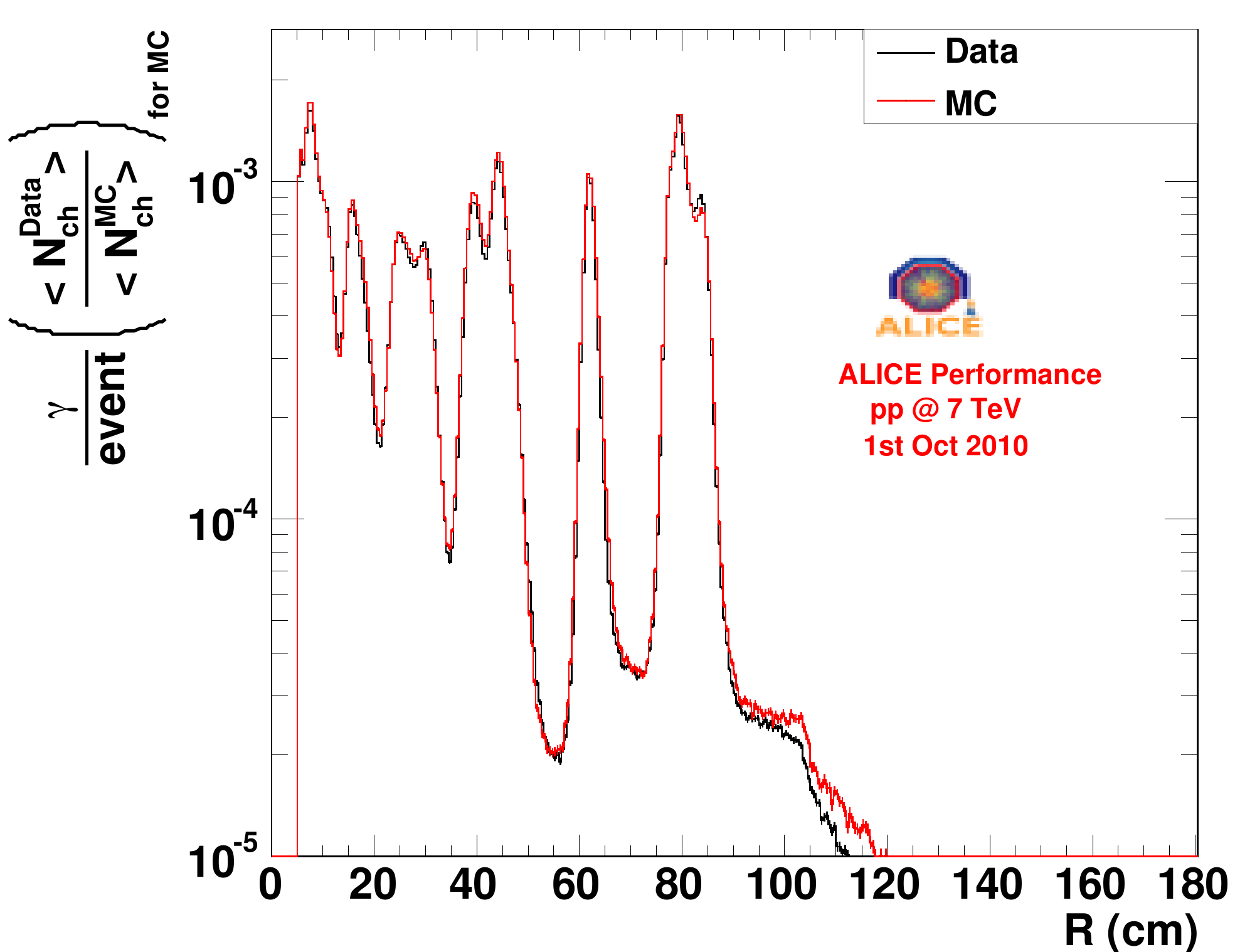}  
  	\end{minipage}
	\hspace{0.5cm}
	\begin{minipage}{7.5 cm}
		\hspace{-0.1cm}
		\includegraphics[height=0.2\textheight]{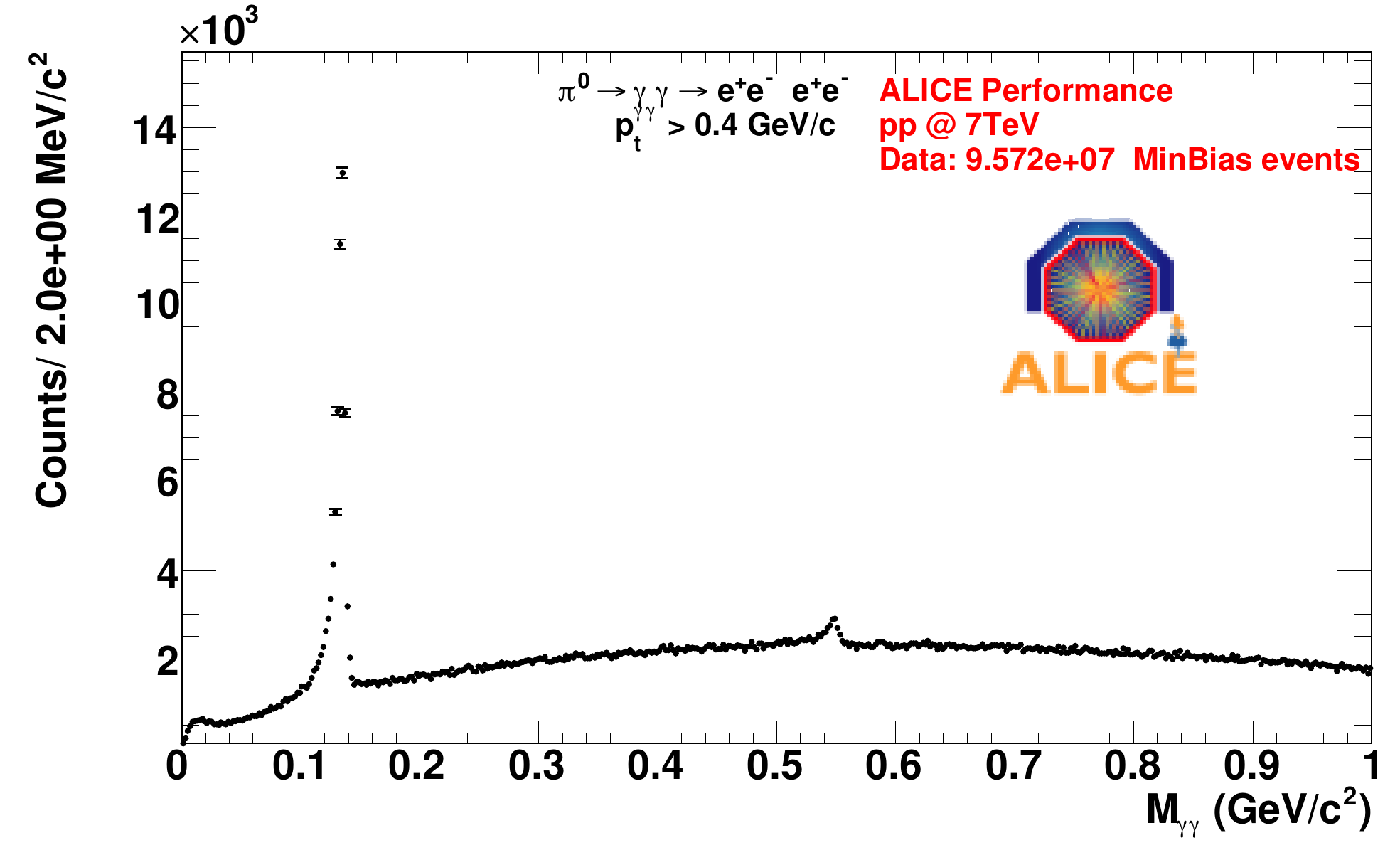}
	\end{minipage}	
	\begin{minipage}{7.5 cm}
		\caption{Comparison of conversion points for data and simulation in R. Both are in good agreement.}
		\label{fig:R}
 	\end{minipage}	
	\hspace{0.5cm}
	\begin{minipage}{7.5 cm}
		\caption{Invariant mass distribution of all photon pairs with $p_{t}^{\gamma\gamma}$ $>$ 0.4 GeV/c reconstructed with the method described in section \ref{sec:PhotonReco}.}
		\label{fig:InvMassDistr}
 	\end{minipage}	
\end{figure}

In this article the results for the reconstruction of the two decays $\pi^{0} \rightarrow \gamma\gamma$ and $\eta \rightarrow \gamma\gamma$ via photon conversions are presented (see figure \ref{fig:eventdisplay}). For both mesons the decay into two photons are the main decay channels with a branching ratio of $98.8 \%$ and $39.3\%$, respectively. A photon with an energy above 1.02 MeV interacts with matter mainly via pair production. It converts in the Coulomb field of a nucleus into an $e^{+}e^{-}$ pair. 
The photon reconstruction is done with an algorithm for displaced vertices. These can be either $K^{0}_{s}$, $\Lambda$ or $\bar \Lambda$ decays, photon conversions or hadronic interactions in the detector material. In order to select photons from this sample, several cuts are applied. These are for example constraints on the energy loss of the track in the TPC to identify electron and positron candidates and on the mass and the pointing direction of the reconstructed photon candidate \cite{alikf}. Furthermore, a condition is used that the tracks should be parallel at the conversion point, otherwise the resolution of the location suffers.
With the conversion method a precise $\gamma$-ray tomography of the ALICE experiment can be obtained with a resolution of the location of the conversion points better than 3 cm in radial direction R, 1.5 cm in beam direction Z and  2.5 mrad in $\varphi$ (see figure \ref{fig:XY}). This provides an important check of the amount of material in the detector and its implementation in GEANT. From detailed comparison of data and simulation the material budget of 0.11 $X_{0}$ is known with an accuracy of $\pm 6\%$ within a photon acceptance of $|\eta|<0.9$ (figure \ref{fig:R}). To determine the conversion probability and the reconstruction efficiency for photons, studies using simulated data have been performed. Within the chosen acceptance the conversion probability is $8.5\%$ for $p_{t}$ $>$ 0.8 GeV/c. The photon reconstruction efficiency, defined as reconstructed photons, that are verified by their Monte Carlo information, divided by the number of converted photons, is found to be around $65\%$ at high $p_{t}$. Since conversion probability and reconstruction efficiency enter squared in the $\pi^{0}$ and $\eta$ meson reconstruction due to the fact of the decay into two photons the efficiency of the method is below $0.3\%$. 
\section{Meson reconstruction}
\label{sec:MesonReco}
\begin{figure}[t]
  \begin{minipage}{7.75 cm}
	\centering
    	\includegraphics[height=0.2\textheight]{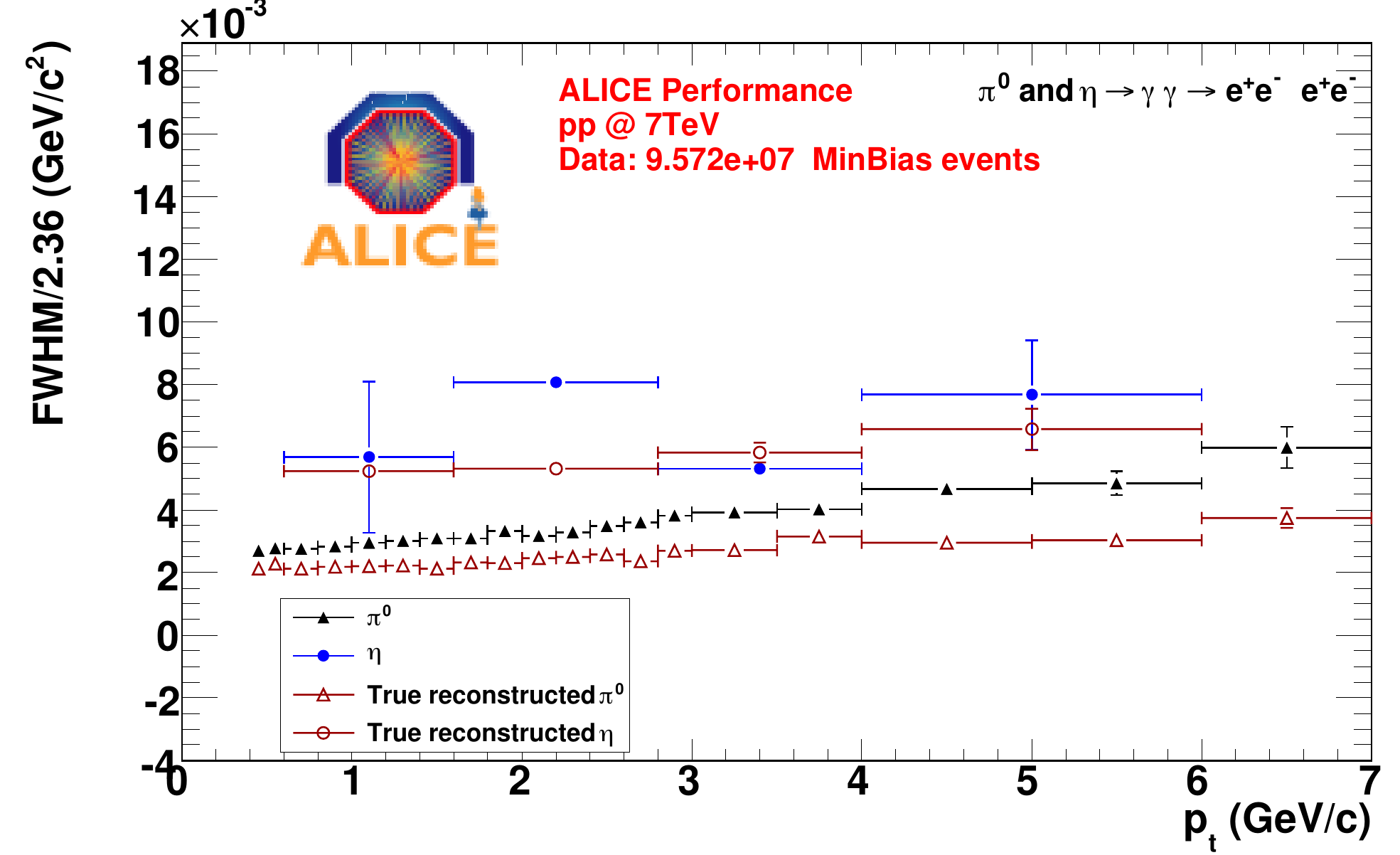}  
    	\caption{Mass resolution (FWHM/2.36) for $\pi^{0}$ and $\eta$ mesons is shown in data and simulation. True reconstructed mesons are verified by their Monte Carlo information.}
    	\label{fig:FWHM}
  \end{minipage}
\hspace{0.5cm}
  \begin{minipage}{7.75 cm}
	\centering
 	\includegraphics[height=0.2\textheight]{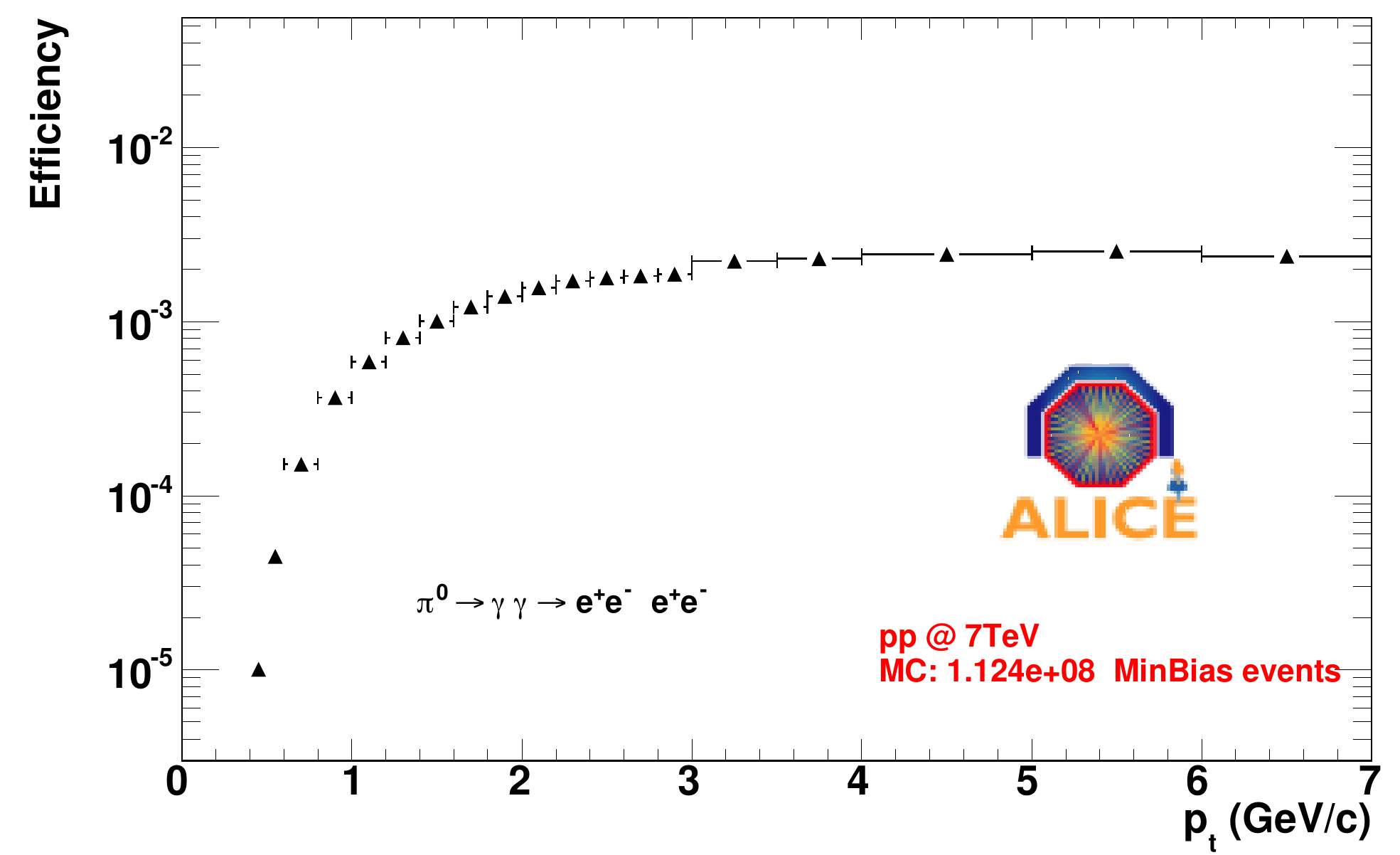}
	\caption{Reconstruction efficiency for $\pi^{0}$ mesons as a function of the transverse momentum.}
	\label{fig:RecoEffPi0}
  \end{minipage}
\end{figure}
In figure \ref{fig:InvMassDistr} the invariant mass distribution of all possible photon pairs with a $p_{t}$ $>$ 0.4 GeV/c, reconstructed via conversions, is shown. Clear peaks at the mass positions of the $\pi^{0}$ and $\eta$ mesons on top of a combinatorial background  can be seen. In order to extract the meson yields the invariant mass distribution is sliced into several $p_{t}$ bins. For the $\pi^{0}$ meson a clear peak is visible in the $p_{t}$ range from 0.4 GeV/c to 7 GeV/c and for the $\eta$ in a range of 0.6 GeV/c to 6 GeV/c, respectively. For the determination of the  combinatorial background a mixed event technique within the same photon multiplicity and $z$ vertex position class is used. The calculated background is normalized for every $p_{t}$ bin in the invariant mass range from 0.17 to 0.3 GeV/$c^{2}$ for the $\pi^{0}$ and 0.58 to 0.62 GeV/$c^{2}$ for the $\eta$, respectively, and subtracted. Afterward the peaks are fitted with a Gaussian function combined with an exponential low energy tail to account for bremsstrahlung and an additional linear part for a possible remaining background: 
\begin{eqnarray*}
y &=& A \cdot \biggl( G + \exp{\biggl( {M_{\gamma\gamma}-M_{\pi^{0}(\eta)}\over{\lambda}}\biggr)}(1-G) \theta(M_{\gamma\gamma}-M_{\pi^{0}(\eta)} )\biggr) + B + C\cdot M_{\gamma\gamma} \\
\mbox{ with  } ~ G&=& exp\biggl(-\frac{1}{2}(\frac{M_{\gamma\gamma}-M_{\pi^{0}(\eta)}}{\sigma})^{2} \biggr) \mbox{ and} ~ \theta (M_{\gamma\gamma} - M_{\pi^{0}(\eta)} ) \mbox{ as the step function}
\end{eqnarray*}
where $G$ is a Gaussian function with the width $\sigma$, amplitude $A$ and center at the invariant mass $M_{\pi^{0}(\eta)}$. $\lambda$ is the inverse slope of the exponential function, that term is switched off by the $\theta$ step function above $M_{\pi^{0}(\eta)}$. $B$ and $C$ are the parameters of the linear function. To determine the meson yields the invariant mass spectra are integrated in fixed invariant mass ranges around the fitted peak position $M_{\pi0(\eta)}$ and the remaining linear background is subtracted. The obtained mass for the $\pi^{0}$ meson is close to the PDG reference of 135 MeV/$c^{2}$, for the $\eta$ it is slightly above. A very good mass resolution (FWHM/2.36) of less than 4 MeV/$c^{2}$ for a transverse momentum below 3 GeV/c and 6 MeV/$c^{2}$ at 7 GeV/c for the $\pi^{0}$ is achieved, the resolution of the $\eta$ peak is below 8 MeV/$c^{2}$ in the whole measured range (see figure \ref{fig:FWHM}). 
\section{Efficiency and acceptance calculations}
\label{sec:Effi}
Using Monte Carlo simulations the geometrical acceptance and the reconstruction efficiency for the $\pi^{0}$ and $\eta$ mesons have been evaluated (see figure \ref{fig:RecoEffPi0}). The reconstruction efficiency for generated $\pi^{0}$ at high $p_{t}$ is slightly below the limit given by the photon reconstruction.  
\section{Results}
\label{sec:Results}
The invariant $\pi^{0}$ ($\eta$) yields are calculated using
\begin{eqnarray*}
\hspace{-0.6cm}
E \frac{d^{3} N^{\pi^{0}(\eta)}}{dp^{3}} =  \frac{d^{3} N^{\pi^{0}(\eta)}}{p_{T} dp_{T} dy d\phi} =  \frac{1}{2\pi} \frac{1}{p_{T}}\frac{d^{2}N^{\pi^0(\eta)}}{dy dp_{T}} = \frac{1}{2\pi} \frac{1}{N_{events}} \frac{1}{p_{T}^{\star}} \frac{1}{\epsilon} \frac{1}{Acc} \frac{1}{BR}\frac{N^{\pi^{0}(\eta)}}{\Delta y \Delta p_{T}}
\end{eqnarray*}
In this formula $N_{events}$ is the number of minimum bias triggered events \cite{trigger}, $\epsilon$ the $\pi^{0}$ ($\eta$) reconstruction efficiency, $Acc$ their geometrical acceptance, BR the branching ratio of $\pi^0$ ($\eta$) mesons to the two $\gamma$ decay channels, $N^{\pi^{0}(\eta)}$ the number of reconstructed $\pi^{0}$ ($\eta$) mesons in a given $\Delta y$ and $\Delta p_{T}$ bin and $p_{T}^{\star}$ the transverse momentum at the bin center. 
\begin{figure}[t!]
 	\begin{minipage}{7.75 cm}
		\includegraphics[width=\textwidth]{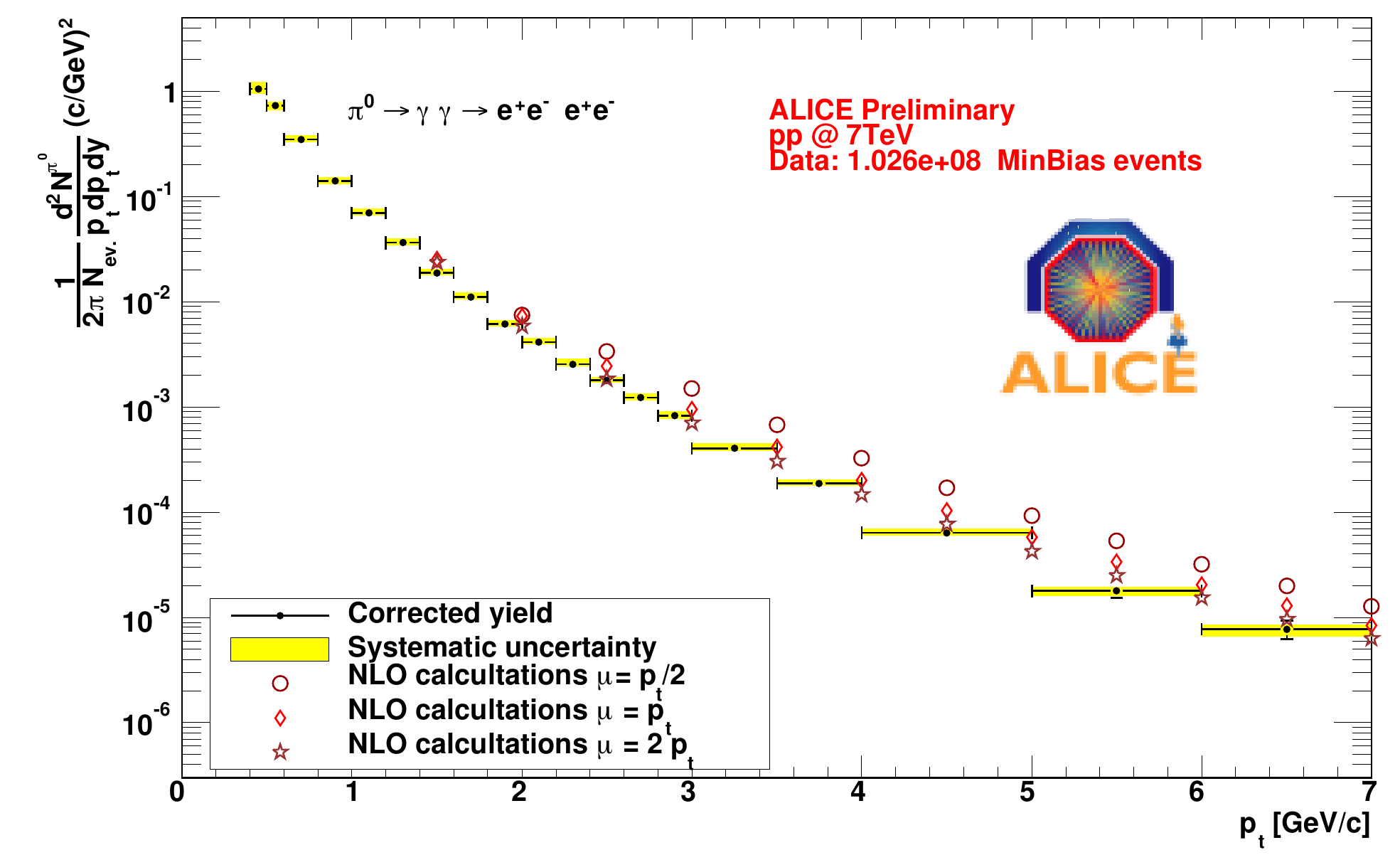}
		\caption{Fully corrected $\pi^{0}$ transverse momentum spectrum. Yellow boxes are the systematic errors of the analysis. An additional systematic error of about 10$\%$ from the overall normalization is not included.}
		\label{fig:corr_spectrum}
 	\end{minipage}
	\hspace{0.25cm}
	\begin{minipage}{7.75 cm}
		\includegraphics[width=0.98\textwidth]{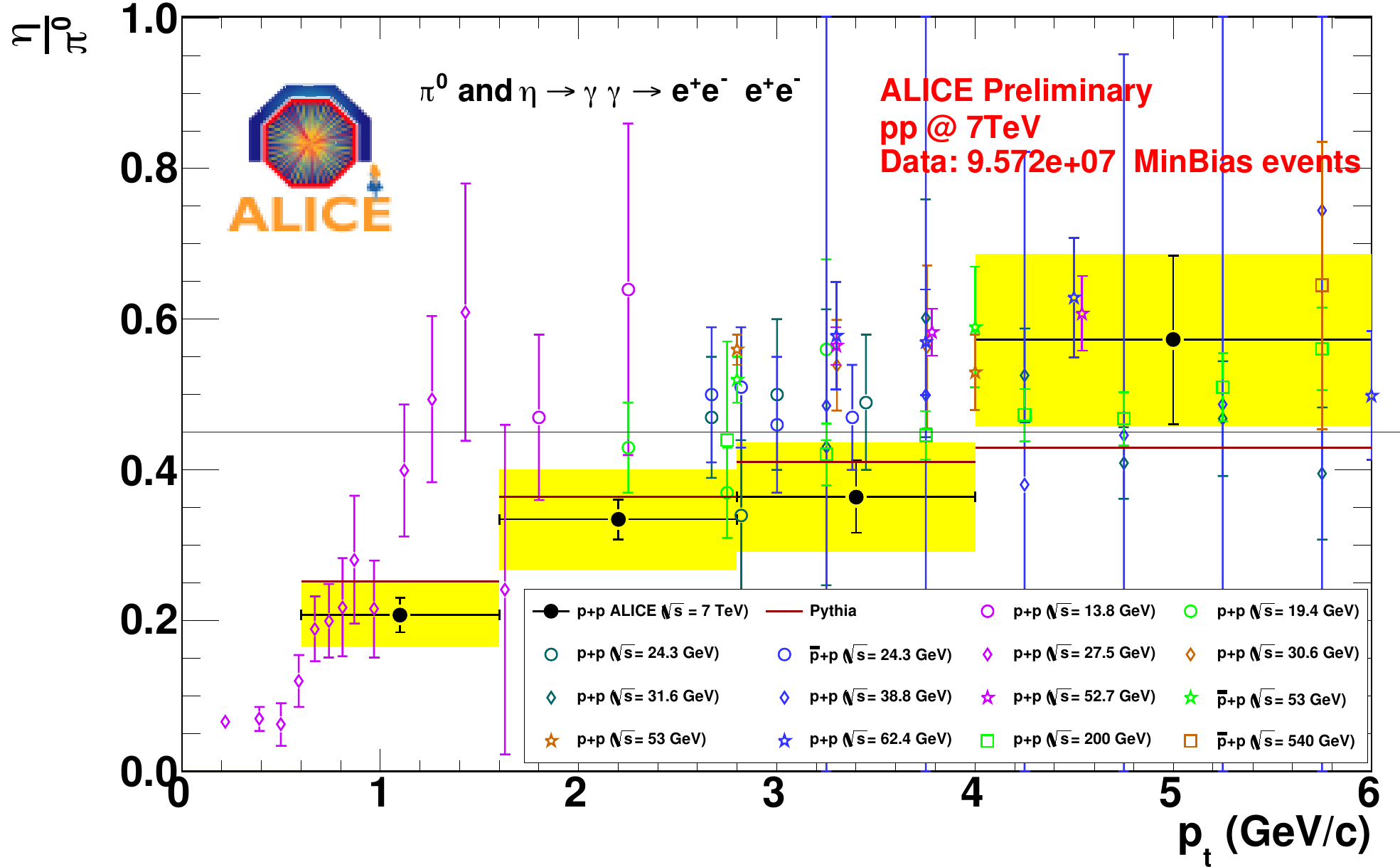}
		\caption{Measured $\eta$/$\pi^{0}$ ratio vs. transverse momentum at $\sqrt{s}$ = 7 TeV. Results are compared to Pythia expectations and world data \cite{world}.}
		\label{fig:etapi0ratio}
 	\end{minipage}	
\end{figure}
In figure \ref{fig:corr_spectrum} the fully corrected $\pi^{0}$ transverse momentum spectrum measured in pp collisions at 7 TeV is presented. It is in agreement with pQCD NLO calculations \cite{NLO1,NLO2,NLO3,NLO4}. For the comparison with the data the NLO calculation was divided by an assumed inelastic pp cross section of $\sigma_{pp}$ of 70 mb. The $\eta$/$\pi^{0}$ ratio is measured in a $p_{t}$ range of 0.6 - 6 GeV/c (see figure \ref{fig:etapi0ratio}). The ratio is in agreement with Pythia predictions and with the world data measured in hadron-hadron collisions \cite{world}. To evaluate the systematic errors different components have been taken into account. These are contributions from the material budget (at present the dominant one) and the signal extraction itself like the minimum $p_{t}$ of the single electron/positron, $dE/dx$ of the electron/positron, $\chi^{2}$ for the photon reconstruction and fraction of the number of reconstructed clusters to the number of detectable clusters for track reconstruction in the TPC. An additional systematic uncertainty on the overall normalization of about 10$\%$ is not included.

\section{Summary and outlook}
\label{sec:Summary}
Photons can be measured in ALICE using reconstructed $e^{+}e^{-}$ pairs from photon conversions with a high precision. Using these converted photons also the reconstruction of $\pi^{0}$ and $\eta$ mesons can be done. The measurement is possible for the $\pi^{0}$ down to very low $p_{t}$ of 0.4 GeV/c with a very high significance and down to 0.6 GeV/c for the $\eta$ meson. For the $\pi^{0}$ meson a preliminary transverse momentum spectrum has been presented. It is in agreement with NLO pQCD calculations. Furthermore the $\eta/\pi^{0}$ ratio is measured in pp collisions at $\sqrt{s}$ = 7 TeV in the $p_{t}$ range from 0.6 - 6 GeV/c. It is in agreement with Pythia expectations and with the world data.


\bibliographystyle{elsarticle-num}



\end{document}